\documentclass{ws-procs975x65}

\begin{document}
%\graphicspath{{fig/}}

\title{HYPERON-QUARK MIXED PHASE IN DENSE MATTER}

\author{TOSHIKI MARUYAMA$^*$ and SATOSHI CHIBA}
\address{Advanced Science Research Center, Japan Atomic Energy Agency, \\
Tokai, Ibaraki 319-1195, Japan\\
$^*$E-mail: maruyama.toshiki@jaea.go.jp}
\author{HANS-JOSEF SCHULZE}
\address{INFN Sezione di Catania, Via Santa Sofia 64, Catania I-95123, Italy}
\author{TOSHITAKA TATSUMI}
\address{Department of Physics, Kyoto University, Kyoto 606-8502, Japan}

\begin{abstract}
We investigate the properties of the hadron-quark mixed phase in compact stars
using a Brueckner-Hartree-Fock framework for hadronic matter and the MIT bag model for quark matter.
We find that the equation of state of the mixed phase is similar to that given
by the Maxwell construction. 
The composition of the mixed phase, however, is very different
from that of the Maxwell construction;
in particular, hyperons are completely suppressed.
\end{abstract}


\bodymatter
\def\sgm{\Sigma^-}

%\begin{table}
%\tbl{Macros available for tables/figures.}
%{\begin{tabular}{@{}ll@{}}
%\toprule
%Environment name&Purpose\\
%\colrule
%{\tt figure} & Figures\\
%{\tt sidewaysfigure} & Landscape figures\\
%{\tt table} & Tables\\
%{\tt sidewaystable} & Landscape tables\\ \colrule
%Horizantal rules & Purpose\\ \colrule
%{\tt $\backslash$toprule} & One rule at the top\\
%{\tt $\backslash$colrule} & One rule separating column heads from\\ & data cells\\
%{\tt $\backslash$botrule} & One bottom rule\\
%{\tt $\backslash$Hline} & One thick rule at the top and bottom of\\
%& the tables with multiple column heads\\ \botrule
%\end{tabular}}
%\end{table}

\def\tsurf{\sigma}
\def\vc{V_{\rm C}}
\def\rv{{\bf r}}

\section{Introduction}

%The theoretical understanding of neutron star (NS) structure requires
%the knowledge of the equation of state (EOS) of highly compressed cold
%baryonic matter, up to densities of about ten times normal nuclear density,
%$\rho_0 \approx 0.17\;\rm fm^{-3}$ \cite{ns}.
%In such an extreme environment, the appearance of ``exotic'' components
%of matter, such as hyperons, meson condensates, and quark matter (QM),
%is expected \cite{nshyp}.

Matter in neutron stars (NS) has a 
variety of density and chemical component due to the presence of gravity.
At the crust of neutron stars,
there exists a region where the density is lower than the normal nuclear
density, $\rho_0\approx 0.17\;\rm fm^{-3}$
over a couple of hundreds meters.
The pressure of such matter is mainly contributed by degenerate electrons,
while baryons are clusterized and have little contribution.
Due to the gravity pressure and density increase
in the inner region
(in fact, the density at the center amounts to several times
%the normal nuclear density
$\rho_0$).
Cold catalyzed matter consists of neutrons and the equal number of protons
and electrons under chemical equilibrium.
Since the kinetic energy of degenerate electrons is much higher than
that of baryons, the electron fraction (or the proton one) decreases
with increase of density and thus neutrons become the main component
and drip out of the clusters.
In this way baryons come to contribute to the pressure as well as electrons.
At a certain density, other components such as hyperons and strange
mesons may emerge.
At even higher density, hadron-quark deconfinement transition may occur
and quarks in hadrons are liberated.

\def\ms{M_{\odot}}
\def\bc{B=100\;\rm MeV/fm^{-3}}

It is well known that hyperons appear at several times 
$\rho_0$ and lead to a strong softening of the EOS
with a consequent substantial reduction of the maximum neutron star mass. 
Actually the microscopic Brueckner-Hartree-Fock approach
gives much lower masses than current observation values of $\sim 1.4M_\odot$. 

On the other hand, the hadron-quark deconfinement transition is believed 
to occur in hot and/or high-density matter.
Taking EOS of quark matter within the MIT bag model, 
the maximum mass can increase to the Chandrasekhar limit once 
the deconfinement transition occurs in hyperon matter \cite{hypns,bal}.

The deconfinement transition from hadron to quark phase may 
occur as a first-order phase transition.
There, a hadron-quark mixed phase should appear, 
where charge density as well as baryon number density is no more uniform. 
Owing to the interplay of the Coulomb interaction 
and the surface tension, the mixed phase can have exotic
shapes called pasta structures \cite{mar}.

Generally, the appearance of mixed phase in matter results in a softening of the EOS.
The bulk Gibbs calculation (BG) of the mixed phase, 
without the effects of the Coulomb interaction and surface tension, 
leads to a broad density region of the mixed phase (MP) \cite{gle92}. % and the softening of EOS.
However, if one takes into account the geometrical structures in the
mixed phase and applies the Gibbs conditions, one may find that MP is
considerably limited and thereby the EOS approaches to the one given by 
the Maxwell construction (MC) \cite{mar,emaru1,maruKaon}.

In this report we explore the EOS and the structure of the mixed phase during the
hyperon-quark transition, 
properly taking account of the Gibbs conditions together with the pasta structures.

\section{Numerical Calculation}

The numerical procedure to determine the EOS and the
geometrical structure of the MP is similar to that 
explained in detail in Ref.~\refcite{mar}.
We employ a Wigner-Seitz approximation in which
the whole space is divided into equivalent Wigner-Seitz 
cells with a given geometrical symmetry,
sphere for three dimension (3D), cylinder for 2D, and slab for 1D.
A lump portion made of one phase is embedded in the other phase and thus 
the quark and hadron phases are spatially separated in each cell.
A sharp boundary is assumed between the two phases and the surface energy
is taken into account in terms of a surface-tension parameter $\tsurf$.
The energy density of the mixed phase is thus written as
\begin{equation}
 \epsilon = {1\over {V_W}} \left[ 
 {\int}_{V_H} d^3 r \epsilon_H({\rv})+
 {\int}_{V_Q} d^3 r \epsilon_Q({\rv})+
 {\int}_{V_W} d^3 r \left( \epsilon_e({\rv}) + {(\nabla \vc({\rv}))^2\over 8\pi e^2} \right)
 + \tsurf S \right] \:,
\end{equation}
where the volume of the Wigner-Seitz cell $V_W$ is the sum of 
those of hadron and quark phases $V_H$ and $V_Q$,
$S$ the quark-hadron interface area.
$\epsilon_H$, $\epsilon_Q$ and $\epsilon_e$ 
are energy densities of hadrons, quarks and electrons,
%$\epsilon_e$ indicates the kinetic energy density of electron.
%The energy densities $\epsilon_H$, $\epsilon_Q$ and $\epsilon_e$ are 
which are 
%$\rv$-dependent since they are 
functions of
local densities $\rho_a(\rv)$ ($a=n,p,\Lambda,\Sigma^-,u,d,s,e$). 
The Coulomb potential $\vc$ is obtained by solving the Poisson equation.
For a given density $\rho_B$, the optimum dimensionality of the cell,
the cell size $R_W$, the lump size $R$,
and the density profile of each component
are searched for to give the minimum energy density.
%The structure of the MP changes from quark droplet to quark slab 
%to hadron tube to hadron bubble with increasing baryon density.
%
%The surface tension of the hadron-quark interface is poorly known, 
%but some theoretical estimates based on the MIT bag model 
%for strangelets \cite{jaf} and
%lattice gauge simulations at finite temperature \cite{latt} suggest
%a range of $\tsurf \approx 10$--$100\;\rm MeV\!/fm^2$.
We employ $\tsurf=40\;\rm MeV\!/fm^2$ 
in the present study. %and discuss the effects of its variation.

To calculate $\epsilon_H$ in the hadron phase,
we use the Thomas-Fermi approximation for the kinetic energy density.
The potential-energy density is 
calculated by the nonrelativistic BHF approach \cite{hypns}
based on microscopic
NN and NY potentials.
\begin{eqnarray}
 \epsilon_H \!&=& \!\!\!\!
 \sum_{i=n,p,\Lambda,\Sigma^-} 
% \int_0^{k_F^{(B)}}{dk\,k^2\over\pi^2} 
 \sum_{k<k_F^{(i)}}
 \left[ T_i(k) + {1\over2} U_i(k) \right] \:,
\end{eqnarray}
\begin{eqnarray}
 U_i(k) &=& 
 \sum_{j=n,p,\Lambda,\Sigma^-} U_i^{(j)}(k)\\
  U_i^{(j)}(k) &=& 
  \!\!\! \sum_{k'<k_F^{(j)}} \!\!\!
  {\rm Re} \big\langle k k' \big| G_{(ij)(ij)}\big[E_{(ij)}(k,k')\big] 
  \big| k k' \big\rangle ,\\
 G_{ab}[W] &=& V_{ab} + \sum_c \sum_{p,p'} V_{ac} \big|pp'\big\rangle 
 {Q_c \over W - E_c +i\epsilon} 
  \big\langle pp'\big| G_{cb}[W] . 
\end{eqnarray}
The interaction parameters are chosen to reproduce the scattering phase shifts.
Nucleonic three-body forces are included in order to (slightly) shift
the saturation point of purely nucleonic matter to the empirical value.

For the quark phase,
we use the MIT bag model with 
massless $u$ and $d$ quarks and massive $s$ quark with $m_s= 150$ MeV.
The energy density $\epsilon_Q$ consists of the kinetic term by the Thomas-Fermi approximation,
the leading-order one-gluon-exchange term \cite{jaf}
proportional to the QCD fine structure constant $\alpha_s$, 
and the bag constant $B$.
% which is the energy-density difference between 
%the perturbative vacuum and the true vacuum.
%Demanding that the quark EOS crosses the hadronic EOS at
%reasonable density, 
Here we use $B=100$ $\rm MeV/fm^3$ and
$\alpha_s=0$ to get the quark EOS which crosses the hadronic one
at an appropriate baryon density.

\section{Hadron-Quark Mixed Phase}

\begin{figure*}%[b]%.............................................................
%\begin{minipage}{0.48\textwidth}
%\includegraphics[width=0.48\textwidth]{ProfQH3D04e}
%\includegraphics[width=0.90\textwidth]{ProfQH3D04e}
\includegraphics[width=1.00\textwidth]{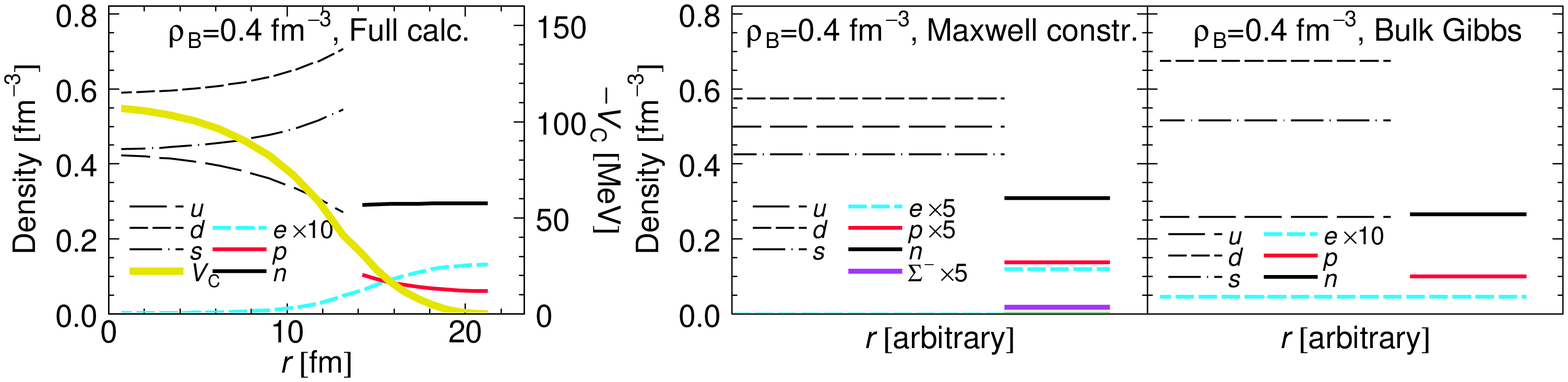}
\caption{
Left:
Density profiles %(thin black lines) 
and Coulomb potential $\vc$ %(thick gray line)
within a 3D (quark droplet) Wigner-Seitz cell
of the MP at $\rho_B=0.4$ fm$^{-3}$.
The cell radius and the droplet radius are $R_W=26.7$ fm
and $R=17.3$ fm, respectively.
Center:
Same as the left panel for MC case.
The radius $r$ is in arbitrary unit since
there is no specific size.
Right:
The case of BG calculation.
}
\label{figProf}
%\end{minipage}
\end{figure*}%..................................................................

Figure~\ref{figProf} illustrates an example of 
the density profile in a 3D cell. % for $\rho_B=0.4$ fm$^{-3}$.
One can see the non-uniform density distribution of each particle species
together with the finite Coulomb potential; charged particle
distributions are rearranged to screen the Coulomb potential.
The quark phase is negatively charged, so that 
$d$ and $s$ quarks are repelled to the phase boundary, 
while $u$ quarks gather at the center.
The protons in the hadron phase are attracted by the negatively charged 
quark phase, while the electrons are localized to the hadron phase.
This density rearrangement of charged particles 
causes the screening of the Coulomb interaction between phases.

In the center and right panels, compared are the cases of MC and BG.
MC assumes the local charge neutrality,
while the BG does not.
One can see that the local charge neutrality in the full calculation 
lies between two cases.
The localization of electrons which is one of the charge screening effects, 
reduces the local charge density.
But the local charge density remains still finite to some extent.

\begin{figure}%[t]%.............................................................
%\hspace{\fill}
%\begin{minipage}{0.48\textwidth}
%\includegraphics[width=0.48\textwidth]{EOS-EP-BG-alp0tau40}
%\includegraphics[width=0.48\textwidth]{EOS-E-alp0tau40BG}
\psfig{width=0.48\textwidth,file=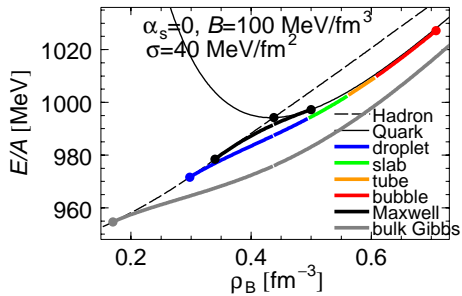}
\caption{
EOS of the MP (thick curves)
in comparison with pure hadron and quark phases (thin curves).
%The upper panel shows the energy per baryon $E/A$ 
%and the lower panel the energy difference between mixed and
hadron ($\rho_B<0.44$ fm$^{-3}$)
or quark ($\rho_B>0.44$ fm$^{-3}$) phases.
Each segment of the MP is chosen by minimizing the energy.
}
\label{figEOS}
%\end{minipage}
\end{figure}%..................................................................

Figure~\ref{figEOS} compares the resulting EOS 
%energy per baryon of 
%the hadron-quark MP 
with that of the pure hadron and quark phases.
%over the relevant range of baryon density. 
The thick black curve indicates the case of the MC,
while the colored lines indicate the MP
in its various geometric realizations
starting %at $\rho_B=0.326$ fm$^{-3}$ 
with a quark droplet structure
and terminating %at $\rho_B=0.666$ fm$^{-3}$ 
with a bubble structure.
Note that the charge screening effect always tends to make matter locally
charge neutral to save the Coulomb interaction energy. Hence,  
combined with the surface tension, it 
%destabilizes the non-uniform structures to limit the MP \cite{mar}.
makes the non-uniform structures mechanically less stable
and limits the density region of the MP \cite{mar}.
Consequently the energy of the MP is only slightly lower than that of the MC. 

However, the structure and the composition of the MP
are very different from those of the MC, 
\begin{figure}%[t]%............................................................
\includegraphics[width=0.80\textwidth]{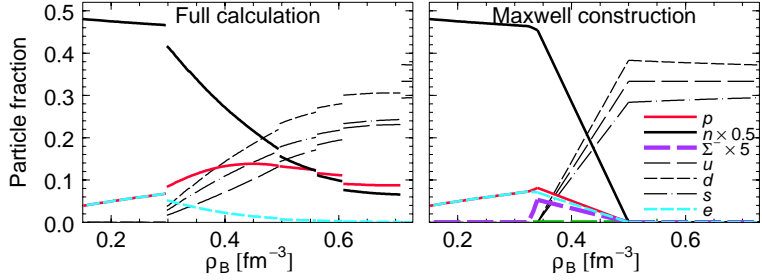}
\caption{
Particle fractions 
in the MP by the full calculation (left panel) and
the MC (right panel).
%In the MC the phase transition occurs between the pure phases with
%$\rho_H=0.34$ fm$^{-3}$ and $\rho_Q=0.50\;\rm fm^{-3}$.
}
\label{figRatio}
%\end{minipage}
\end{figure}%..................................................................
which is demonstrated in Fig.~\ref{figRatio}, where we compare the 
particle fractions as a function of baryon density in the full calculation (left panel)
and the MC (right panel).
One can see that the compositions are very different in two cases.
%the MP by the full calculation lying in between the extreme cases of Bulk Gibbs and MC.
In particular, a relevant hyperon ($\Sigma^-$) fraction is only present in the MC.

\begin{figure}[t]%............................................................
\includegraphics[width=0.80\textwidth]{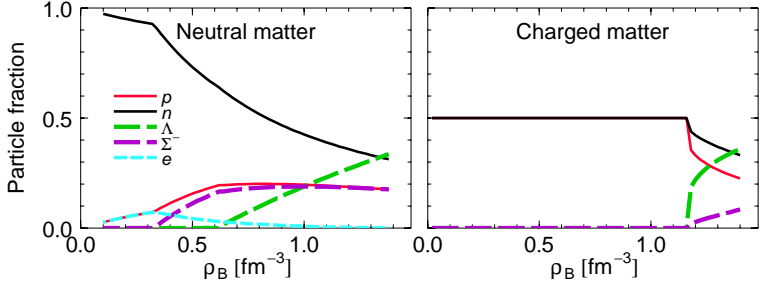}
\caption{
Upper panel:
Particle fractions of neutral matter with electrons (corresponding to neutron star matter).
Lower panel:
The same quantity for charged matter without electrons,
the low-density part of which corresponds to symmetric nuclear matter.
%Both cases require beta-equilibrium condition.
Both cases require chemical-equilibrium condition.
}
\label{figRatioUnif}
\end{figure}%.................................................................

The suppression of hyperon mixture in the MP
is due to the absence of the
charge-neutrality condition in each phase.
As shown in Fig.\ \ref{figRatioUnif}, 
hyperons ($\Sigma^-$) appear in charge-neutral hadronic matter 
at low density ($0.34\;\rm fm^{-3}$)
to reduce the Fermi energies of electron and neutron.
Without the charge-neutrality condition, on the other hand, 
there appears symmetric nuclear matter at lower density
and hyperons will be mixed above $1.15\;\rm fm^{-3}$
due to the large hyperon masses.
%In the MP the hadron phase can be
%positively charged due to the presence of the negatively charged quark phase.
The MP has positively charged hadron phase and negative quark phase
though the Coulomb screening effect diminishes the local charge density.
This brings the feature of symmetric nuclear matter into
the hadron phase and, consequently, the mixture of hyperons is suppressed.\cite{maruhyp}

\section{Neutron Star Structure}

Having the EOS comprising hadronic, mixed, and quark phase
in the form $P(\epsilon)$,
the equilibrium configurations of static NS are obtained
in the standard way
by solving the Tolman-Oppenheimer-Volkoff (TOV) equations \cite{ns} for 
the pressure $P(r)$ and the enclosed mass $m(r)$,
\begin{eqnarray}
  {dP\over dr} &=& -{ G m \epsilon \over r^2 } \,
%\nonumber\\ && \times
  {  \left( 1 + {P / \epsilon} \right) 
  \left( 1 + {4\pi r^3 P / m} \right) 
  \over
  1 - {2G m/ r} } \:,\qquad
\\
  {dm \over dr} &=& 4 \pi r^2 \epsilon \:,
\end{eqnarray}
being $G$ the gravitational constant. 
Starting with a central mass density $\epsilon(r=0) \equiv \epsilon_c$,  
one integrates out until the surface density equals the one of iron.
This gives the stellar radius $R$ and its gravitational mass $M=m(R)$.
For the description of the NS crust, we have joined 
the hadronic EOS with the ones by 
Negele and Vautherin \cite{nv} in the medium-density regime, and the ones   
by Feynman-Metropolis-Teller \cite{fey} 
and Baym-Pethick-Sutherland \cite{baym} for the outer crust.

%-----------------------------------------------------------------------------
%\section{Results}
\label{s:res}

\begin{figure}[t]%............................................................
\includegraphics[width=0.48\textwidth]{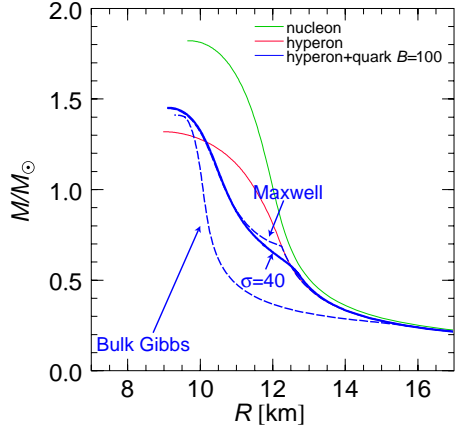}
\caption{
(Color online)
Neutron star mass-radius relations for different EOS 
and three different hadron-quark phase transition constructions.
For the hybrid stars (blue and black curves),
the dashed lines indicate the Maxwell (upper curves)
or bulk Gibbs (lower curves) constructions
and the solid lines the mixed phase of the full calculation.
}
\label{f:mr}
\end{figure}%.................................................................

\begin{figure*}[t]%...........................................................
\includegraphics[width=0.98\textwidth]{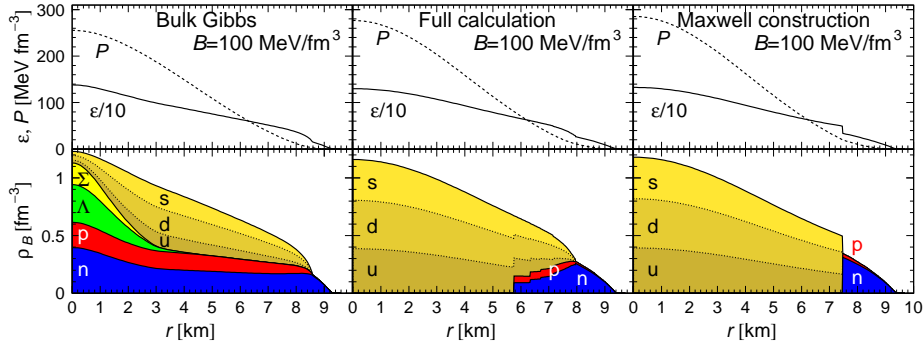}
\caption{
(Color online)
Internal structure of a $1.4\,\ms$ neutron star
obtained with three different phase transition constructions.
The upper panels show total energy density and pressure
and the lower panels the overall particle fractions as functions
of the radial coordinate of the star,
using the bulk Gibbs calculation (left panel), 
the mixed phase of the full calculation with 
$\sigma=40\;\rm MeV\!/fm^2$ (central panel),
and the Maxwell construction (right panel).
In all cases $\alpha_s=0$ and $\bc$ are used.
}
\label{f:xr1}
\end{figure*}%................................................................

Fig.~\ref{f:mr} compares the mass-radius relations obtained with the 
different models.
The purely nucleonic EOS (green curve) yields a maximum NS mass
of about $1.82\,\ms$, which is reduced to $1.32\,\ms$ when allowing
for the presence of hyperons (red curve).
This feature has been shown to be fairly independent of the nucleonic 
and hyperonic EOS that are used \cite{hypns2}.
The canonical NS with mass of about $1.4\,\ms$ 
can therefore not be purely hadronic stars in our approach.
In fact, the inclusion of quark matter augments the maximum mass 
of hybrid stars to about $1.5\,\ms$:

%More precisely, we compare in the figure results obtained with the two quark EOS
%$\bc$ (blue curves) and $\beff$ (black curves), and involving the 
%different phase transition constructions Bulk, Mixed, and Maxwell.
In general, the Maxwell construction leads to a kink in the $M(R)$ relation,
because the transition from a hadronic to a hybrid star occurs suddenly,
involving a discontinuous increase of the central density when the quark
phase onsets in the core of a star.
The bulk Gibbs calculation yields smooth mass-radius relations 
involving a continuous transition from a hadronic to a hybrid star 
beginning at rather low central density corresponding to very low NS mass.

The MP construction by the full calculation lies between the two extreme cases,
and with our choice of $\sigma=40\;\rm MeV\!/fm^2$
it is rather close to the Maxwell construction,
smoothing out the kink of the hadron-hybrid star transition.
This transition occurs generally at a fairly low NS mass,
even below the natural minimum mass limit due to the 
formation via a protoneutron star \cite{proto}
and is thus an unobservable feature.

%%On the contrary, the maximum mass is hardly affected by the
%type of phase transition:
%For the $\beff$ model the maximum mass is $1.52\,\ms$, 
%practically independent of the kind of phase transition, 
%whereas for the $\bc$ model there is a slight variation of 
%$M=1.45,1.45,1.41\,\ms$ for the Maxwell, mixed, and bulk construction,
%respectively.

Whereas the maximum masses are practically independent of the phase 
transition construction, there are evidently large differences for the internal
composition of the star.
This is illustrated in Fig.~\ref{f:xr1} %and \ref{f:xr2}, 
which show the 
total energy density, pressure, and
particle fractions as a function of the radial coordinate for a
$1.4\,\ms$ NS.
One observes with the bulk Gibbs construction (left panels)
a coexistence of hadrons and quarks 
% throughout the whole interior of the star, 
in a significant range of the star,
whereas with the MC (right panels)
an abrupt transition involving a discontinuous 
jump of energy and baryon density occurs at a distance 
$r\approx7.5\;\rm km$ from the center of the star.
The small contamination with $\sgm$ hyperons in the hadronic phase
is not visible on the scale chosen.
The MP with the full calculation (central panels) lie between 
the two extreme cases, 
%and the hadrons disappear gradually when
%hadrons and quarks coexisting in the range $r\approx5.7$ -- $8.0\;\rm km$.
hadrons and quarks coexisting in a smaller range than in the bulk Gibbs cases.
%In both latter cases the pure quark matter core has a higher central pressure
%and baryon number and energy densities 
%than the mixed core of the first case.

%Note that the $\beff$ calculations (Fig.~\ref{f:xr2}) yield a
%very narrow range of the MP compared to the $\bc$ case (Fig.~\ref{f:xr1}).
%At a given baryon density
%the energy density and pressure of quark matter with $\beff$ are much 
%lower than with $\bc$ (see Fig.~\ref{figParam})
%%and therefore the system jumps immediately to a higher 
%baryon (quark) density in the MC.
%Since the same tendency is seen for the bulk Gibbs calculation,
%the difference between the MC and the bulk Gibbs EOS is rather small 
%%%%%in the $\beff$ case.
%Consequently within the full calculation the density range for 
%the phase coexistence is overcome
%very quickly and the result is very close to the MC;
%in fact the mixed phase only exists in the narrow interval 
%%$r=7.56,\ldots,7.71\;\rm km$,
%$r=7.56$--$7.71\;\rm km$,
%hardly visible in the plot.

%The two quark models thus illustrate two extreme cases of the same generic
%%phenomenon, namely that the full MP calculation 
%turns out very close to the MC.

\section{Summary}

In this article we have studied the properties of the mixed phase
in the quark deconfinement transition in hyperonic matter, and their
influence on compact star structure.
%the influence of different constructions
%for the hadron-quark phase transition in beta-stable matter
%on neutron star observables...
The hyperonic EOS given by the BHF approach with realistic hadronic
interactions is so soft that the transition density becomes
very low if one uses the MIT bag model for the quark EOS.

The hyperon-quark mixed phase was consistently treated with the
basic thermodynamical requirement due to the Gibbs conditions.
We have seen that the resultant EOS
is little different from the one given by the Maxwell construction.
This is because the finite-size effects, the surface tension,
and the Coulomb interaction tend to diminish the available density region
through the mechanical instability,
as has also been suggested in previous articles \cite{vos,mixtat}.

For the bulk properties of compact stars, such as mass or radius,
our EOS gives similar results as those given by the Maxwell construction.
The maximum mass of a hybrid star is around $1.5\,M_\odot$, larger than
that of the purely hyperonic star, $\approx1.3\,M_\odot$.
Hence we may conclude that a hybrid star is still consistent with the
canonical NS mass of $1.4\,M_\odot$,
while the masses of purely hyperonic stars lie below it.

On the other hand, the internal structure of the mixed phase is very
different; e.g., the charge density as well as the baryon number density
are nonuniform in the mixed phase.
We have also seen that the hyperon number fraction
is %completely
suppressed in the mixed phase due to the relaxation
of the charge-neutrality condition,
while it is always finite in the Maxwell construction.
This has important consequences for the elementary processes inside compact
stars.
For example, coherent scattering of neutrinos off lumps in the
mixed phase may enhance the neutrino opacity \cite{red}.
Also, the absence of hyperons prevents a fast cooling mechanism by way of
the hyperon Urca processes \cite{pet}.
These results directly modify the thermal evolution of compact stars.

%\bibliographystyle{ws-procs975x65}
%\bibliography{ws-pro-sample}

\begin{thebibliography}{00}

%\bibitem{nshyp}
% PHASES OF DENSE MATTER IN NEUTRON STARS
% H.~Heiselberg and M.~Hjorth-Jensen,
%  {\em Phys. Rep.} {\bf 328}, 237 (2000).


\bibitem{hypns}
 M.~Baldo, G.~F.~Burgio, and H.-J.~Schulze, {\em Phys. Rev. C} {\bf 58}, 3688 (1998); 
 {\em Phys. Rev.} {\bf C61}, 055801 (2000).

\bibitem{bal}
M.~Baldo, G.~F.~Burgio and H.-J.~Schulze, {\em Phys.\ Rev.} {\bf C61}, 055801 (2000); 
O.E.~Nicotra, M.~Baldo, G.F.~Burgio and H.-J.~Schulze, {\em A \& A} {\bf 451}, 213 (2006).

\bibitem{mar}
For a review, T.~Maruyama, T.~Tatsumi, T.~Endo and S.~Chiba, 
{\em Recent Res.\ Devel.\ Phys.} {\bf 7}, 1 (2006).

\bibitem{gle92}
N.~K.~Glendenning, {\em Phys. Rev.} {\bf D46}, 1274 (1992);
%N.~K.~Glendenning, 
{\em Phys. Rep.} {\bf 342}, 393 (2001).


\bibitem{emaru1}
T.~Endo, T.~Maruyama, S.~Chiba and T.~Tatsumi, 
{\em Prog. Theor. Phys.} {\bf 115}, 337 (2006).


\bibitem{maruKaon}
T.~Maruyama, T.~Tatsumi, D.~N.~Voskresensky, T.~Tanigawa and S.~Chiba, 
{\em Phys. Rev.} {\bf C73}, 035802 (2006).  


\bibitem{jaf}
 E.~Farhi and R.~L.~Jaffe, {\em Phys. Rev.} {\bf D30}, 2379 (1984);
 M.~S.~Berger and R.~L.~Jaffe, {\em Phys. Rev.} {\bf C35}, 213 (1987).


\bibitem{maruhyp}
T.~Maruyama, S.~Chiba, H.-J.~Schulze and T.~Tatsumi, {\em Phys. Lett.} {\bf B}, inpress;
{\em Phys. Rev.} {\bf D}, inpress.

\bibitem{ns}
%
 S. L. Shapiro and S. A. Teukolsky,
  {\it Black Holes, White Dwarfs and Neutron Stars}
   (Wiley, New York, 1983).

\bibitem{nv}
% NEUTRON STAR MATTER AT SUB-NUCLEAR DENSITIES
 J. W. Negele and D. Vautherin,
 {\em Nucl. Phys.} {\bf A207}, 298 (1973).

\bibitem{fey}
% EQUATIONS OF STATE OF ELEMENTS BASED ON THE GENERALIZED FERMI-THOMAS THEORY
 R. P. Feynman, N. Metropolis, and E. Teller,
 {\em Phys. Rev.} {\bf 75}, 1561 (1949).

\bibitem{baym}
% THE GROUND STATE OF MATTER AT HIGH DENSITIES:
% EQUATION OF STATE AND STELLAR MODELS
 G. Baym, C. Pethick, and P. Sutherland,
 {\em Astrophys. J.} {\bf 170}, 299 (1971).

\bibitem{hypns2}
% MAXIMUM MASS OF NEUTRON STARS
 H.-J. Schulze, A. Polls, A. Ramos, and I. Vida\a~na,
 {\em Phys. Rev.} {\bf C73}, 058801 (2006).

\bibitem{proto}
% PROTONEUTRON STARS WITHIN THE BRUECKNER-BETHE-GOLDSTONE THEORY
 O. E. Nicotra, M. Baldo, G. F. Burgio, and H.-J. Schulze,
 {\em Astron. Astrophys.} {\bf 451}, 213 (2006);
% HYBRID PROTONEUTRON STARS WITH THE MIT BAG MODEL
 {\em Phys. Rev.} {\bf D74}, 123001 (2006).

\bibitem{vos}
% CHARGE SCREENING IN HADRON-QUARK MIXED PHASE
 D. N. Voskresensky, M. Yasuhira, and T. Tatsumi,
 {\em Phys. Lett.} {\bf B541}, 93 (2002);
% CHARGE SCREENING AT FIRST ORDER PHASE TRANSITIONS AND HADRON-QUARK MIXED PHASE
 D. N. Voskresensky, M. Yasuhira, and T. Tatsumi,
 {\em Nucl. Phys.} {\bf A723}, 291 (2003).

\bibitem{mixtat}
% HADRON-QUARK MIXED PHASE IN NEUTRON STARS
 T. Tatsumi, M. Yasuhira, and D. N. Voskresensky,
 {\em Nucl. Phys.} {\bf A718}, 359 (2003);
% NUMERICAL STUDY OF THE HADRON-QUARK MIXED PHASE
 T. Endo, T. Maruyama, S. Chiba, and T. Tatsumi,
{\em  Nucl. Phys.} {\bf A749}, 333 (2005);
% CHARGE SCREENING EFFECT IN THE HADRON-QUARK MIXED PHASE
 %T. Endo, T. Maruyama, S. Chiba, and T. Tatsumi,
 %{\em Prog. Theor. Phys.} {\bf 115}, 337 (2006).

\bibitem{red}
% FIRST ORDER PHASE TRANSITIONS IN NEUTRON STAR MATTER:
% DROPLETS AND COHERENT NEUTRINO SCATTERING
 S. Reddy, G. Bertsch, and M. Prakash,
 {\em Phys. Lett.} {\bf B475}, 1 (2000).

\bibitem{pet}
% RAPID COOLING OF NEUTRON STARS BY HYPERONS AND DELTA ISOBAR
 M. Prakash, M. Prakash, J. M. Lattimer, and C. J. Pethick,
 {\em Astrophys. J.} {\bf 390}, L77 (1992);
 % NEUTRINO INTERACTIONS IN OCTET BARYON MATTER
 T. Tatsumi, T. Takatsuka, and R. Tamagaki,
 {\em Prog. Theor. Phys.} {\bf 110}, 179 (2003);
 %
 T. Takatsuka, S. Nishizaki, Y. Yamamoto, and R. Tamagaki,
{\em Prog. Theor. Phys.} {\bf 115}, in press (2007).

\end{thebibliography}

\end{document}